\documentclass[11pt,english,onecolumn, draftcls]{IEEEtran}
\usepackage[T1]{fontenc}
\usepackage[latin9]{inputenc}
\usepackage{color}
\usepackage{float}
\usepackage{amsmath}
\usepackage{amsthm}
\usepackage{amssymb}
\usepackage{graphicx}

\makeatletter

\floatstyle{ruled}
\newfloat{algorithm}{tbp}{loa}
\providecommand{\algorithmname}{Algorithm}
\floatname{algorithm}{\protect\algorithmname}

\theoremstyle{plain}
\newtheorem{thm}{\protect\theoremname}
\theoremstyle{plain}
\newtheorem{lem}[thm]{\protect\lemmaname}
\theoremstyle{plain}
\newtheorem{prop}[thm]{\protect\propositionname}
\theoremstyle{plain}
\newtheorem{cor}[thm]{\protect\corollaryname}

\makeatother

\usepackage{babel}
\providecommand{\corollaryname}{Corollary}
\providecommand{\lemmaname}{Lemma}
\providecommand{\propositionname}{Proposition}
\providecommand{\theoremname}{Theorem}

\begin{document}

\title{Approximation Algorithms for Minimizing Maximum Sensor Movement\textcolor{black}{{}
for Line Barrier Coverage }in the Plane \textcolor{black}{}\thanks{\textcolor{black}{This research work is supported by Natural Science
Foundation of China \#61300025, Doctoral Fund of Ministry of Education
of China for Young Scholars \#20123514120013, and Australian Research
Discovery Project DP 150104871. }} }

\author{\textcolor{black}{Longkun Guo$^{1}$, Hong Shen$^{2}$}\linebreak{}
\textcolor{black}{{} $^{1}$College of Mathematics and Computer Science,
Fuzhou University, China}\\
\textcolor{black}{{} $^{2}$School of Computer Science, University of
Adelaide, Australia }\\
}
\maketitle
\begin{abstract}
Given a line barrier and a set of mobile sensors distributed in the
plane, the Minimizing Maximum Sensor Movement problem (MMSM) for \textcolor{black}{line
barrier coverage} is to compute relocation positions for the sensors
in the plane such that the barrier is entirely covered by the monitoring
area of the sensors while the maximum relocation movement (distance)
is minimized. Its weaker version, decision MMSM is to determine whether
the barrier can be covered by the sensors within a given relocation
distance bound $D\in\mathbb{Z}^{+}$.

This paper presents three approximation algorithms for decision MMSM.
The first is a simple greedy approach, which runs in time $O(n\log n)$
and achieves a maximum movement $D^{*}+2r_{max}$, where $n$ is the
number of the sensors, $D^{*}$ is the maximum movement of an optimal
solution and $r_{max}$ is the maximum radii of the sensors. The second
and the third algorithms improve the maximum movement to $D^{*}+r_{max}$
, running in time $O(n^{7}L)$ and $O(R^{2}\sqrt{\frac{M}{\log R}})$
by applying linear programming (LP) rounding and maximal matching
tchniques respecitvely, where $R=\sum2r_{i}$, which is $O(n)$ in
practical scenarios of uniform sensing radius for all sensors, and
$M\leq n\max r_{i}$. Applying the above algorithms for $O(\log(d_{max}))$
time in binary search immediately yields solutions to MMSM with the
same performance guarantee. In addition, we also give a factor-2
approximation algorithm which can be used to improve the performance
of the first three algorithms when $r_{max}>D^{*}$. As shown in \cite{dobrev2015complexity},
the 2-D MMSM problem admits no FPTAS as it is strongly NP-complete,
so our algorithms arguably achieve the best possible ratio.
\end{abstract}


\section{Introductions}

Barrier coverage and area coverage are two important problems in applications
of wireless sensor networks. In both two problems, sensors are deployed
in such a way that every point of the target region is monitored by
at least one sensor. For area coverage, the target region is traditionally
a bounded area in the plane; while in the barrier coverage problem
arising from border surveillance for intrusion detection, the target
region are the borders and the goal is to deploy sensors along the
borders such that at least one sensor will detect if any intruder
crosses the border. Unlike area coverage, barrier coverage requires
only to cover every points of the borders, rather than every point
of the area bounded by the border. So barrier coverage uses much fewer
sensors, and hence is more cost-efficient, particularly in practical
large-scale sensor deployment.

To accomplish a barrier coverage, sensors are dispersed along the
borders. However, there may exist gaps after the dispersal, so the
border line might not be completely covered. One approach is to disperse
the sensors in multiple rounds, and guarantee the probability of complete
coverage by the dispersal density of the sensors \cite{yang2010multi,li2012energy}.
The other approach is to acquire some sensors with the ability of
relocation (i.e. mobility), such that after dispersal, the sensors
can move to monitor the gaps on the barrier. In this context, since
the battery of a sensor is limited, a smart relocation scheme is required
to maximize the lifetime of the sensors, and hence ensures a maximum
lifetime of the barrier coverage.

\subsection{Problem Statement}

This paper studies the two-dimensional (2-D) barrier coverage problem
with mobile sensors, in which the barrier is modeled by a line segment,
while the sensors are distributed in the plane initially. The problem
is to compute the relocated positions of the sensors, such that the
barrier will be completely covered while the maximum relocation distance
among all the sensors is minimized.

Formally, we are given a line barrier $[0,M]$ on $x$-axis and a
set of sensors distributed in the Euclidean plane, say $\Gamma=\{1,\,\dots,\,n\}$,
within which sensor $i$ is with a radii $r_{i}$ and a position $(x_{i},\,y_{i})$.
The two dimensional Minimum Maximum Sensor Movement problem (MMSM)
is to compute the minimum $D\in\mathbb{Z}_{0}^{+}$ and a new position
$(x_{i}',\,0)$ for each sensor $i$, such that $\max_{i\in n}\sqrt{(x_{i}-x_{i}')^{2}+(y_{i}-0){}^{2}}\leq D$
and each point on the line barrier is covered by at least one sensor
(i.e. for each point on the line barrier there exists at least a sensor
$i$ within distance $r_{i}$).

The paper finds that, MMSM can be reduced to a discrete version called
DMMSM. In DMMSM, we are given a graph $G=(V,\,E)$, where $V=\{v_{0},\,v_{1},\,\dots,\,v_{M}\}$
and $E=\{e_{i}=(v_{i},\,v_{i+1})\vert i\in V\setminus\{M\}\}$. We
say an edge $e_{j}$ is covered by a set of sensors $\Gamma'\subseteq\Gamma$
if and only if every point on the edge is in the monitoring area of
at least one sensor of $\Gamma'$. The goal of DMMSM is also to compute
a minimum maximum movement $D\in\mathbb{Z}_{0}^{+}$ and the new relocate
position for each sensor, such that every edge of $E$ is covered
by the sensors.

We propose several algorithms that are actually first to solve decision
MMSM and decision DMMSM, which is to determine, for a given the relocation
distance bound $D$, whether the sensors can be relocated within $D$
to cover the line barrier.

\subsection{Related Works}

The MMSM problem in 2D setting was first studied in \cite{dobrev2015complexity},
and shown strongly ${\cal NP}$-complete for sensors with general
sensing radii via a reduction from the 3-partition problem which is
known strongly ${\cal NP}$-complete. Later, an algorithm with a time
complexity of $O(n^{3}\log n)$ has been developed in \cite{li2015minimizing}
for the problem where sensors are with identical sensing radii. In
the same paper, an approximation algorithm with ratio $\frac{y_{max}}{y_{min}}$
for general radii has also been developed, where $y_{max}$ and $y_{min}$
are respectively the maximum and minimum perpendicular relocation
distance from the sensors to the barrier. To the best of our knowledge,
there is no any other non-trivial approximation algorithm for MMSM
with general radii.

Unlike 2-D MMSM, the MMSM problem has been extensively studied and
well understood in 1-D setting, in which the barrier are assumed to
be a line segment on the same line where the sensors are initially
located. Paper \cite{czyzowicz2009minimizing} presented an algorithm
which optimally solves 1D-MMSM for uniform radius and runs in time
$O(n^{2})$, by observing the order preservation property. The time
complexity was improved to $O(n\log n)$ later in paper \cite{chen2013algorithms},
which also gave an $O(n^{2}\log n)$ time algorithm for general radii.
Recently, an $O(n^{2}\log n\log\log n)$ time algorithm has been presented
in \cite{wang2015minimizing} for weighted 1D-MMSM with uniform radii,
in which each sensor has a weight, and the moving cost of a sensor
is its movement times its weight. Moreover, circle/simple polygon
barriers has been studied besides straight lines in \cite{bhattacharya2009optimal},
in which two algorithms has been developed for MMSM, with an $O(n^{3.5}\log n)$
time against cycle barriers and an $O(mn^{3.5}\log n)$ time against
polygon barriers, where $m$ is the number of the edges on the polygon.
The later time complexity was then improved to $O(n^{2.5}\log n)$
in \cite{tan2010new}.

Other problems closely related to MMSM have also been well studied
in previous literature. In 1-D setting, the Min-Sum relocation problem,
to minimize the sum of the relocation distances of all the sensors,
is shown ${\cal NP}$-complete for general radii while solvable in
time $O(n^{2})$ for uniform radii \cite{czyzowicz2010minimizing}.
The Min-Num relocation problem of minimizing the number of sensors
moved, is also proven ${\cal NP}$-complete for general radii and
polynomial solvable for uniform radii \cite{mehrandish2011minimizing}.
Similar to MMSM, where a PTAS has been developed for the Min-Sum relocation
problem against circle/simple polygon barriers \cite{bhattacharya2009optimal},
which was improved by later paper \cite{tan2010new} that gave an
$O(n^{4})$ time exact algorithm.

Paper \cite{bar2013maximizing} studied a more complicated problem
of maximizing the coverage lifetime, in which each mobile sensor is
equipped with limited battery power, and the coverage lifetime is
the time to when the coverage no longer works because of the death
of a sensor. The authors presented parametric search algorithms for
the cases when the sensors have a predetermined order in the barrier
or when sensors are initially located at barrier endpoints. On the
other hand, the same authors present two FPTAS respectively for minimizing
sumed and maximum energy consumption when the radii of the sensors
can be adjusted \cite{bar2015green}. When the sensing radii is fixed,
i.e. unadjustable, the same paper showed the min-sum problem can not
be approximated within $O(n^{c})$ for any constant $c$ under the
assumption of ${\cal P}\neq{\cal NP}$, while the min-max version
is known strongly ${\cal NP}$-complete, as it can be reduced to 2-D
MMSM which is know strongly ${\cal NP}$-complete \cite{dobrev2015complexity}.

Before deployment of mobile sensors, barrier coverage was first considered
deploying stationary sensors \cite{kumar2005barrier} for covering
a closed curve (i.e. a moat), and an elegant algorithm was proposed
by transferring the Min-Sum cost barrier coverage problem to the shortest
path problem. It has then been extensively studied for line based
employment \cite{saipulla2009barrier}, for better local barrier coverage
\cite{chen2010local}, and for using camera sensors \cite{wang2011barrier,ma2012minimum}.
The most recent result \cite{fan2014barrier} studied line barrier
coverage using sensors with adjustable sensing ranges. They show the
problem is polynomial solvable when each sensor can only choose from
a finite set of sensing ranges, and ${\cal NP}$-complete if each
sensor can choose any sensing ranges in a given interval.

\subsection{Our Results and Technique}

In this paper, we present two approximation algorithms for the decision
MMSM problem. The first is a simple greedy approach based on our proposed
sufficient condition of determining whether there exists a feasible
cover for the barrier under the relocation distance bound $D$. If
$D<D^{*}$, the algorithm outputs ``infeasible''; Otherwise, the
algorithm computes new positions for the sensors, resulting a maximum
relocation distance $D+2r_{max}$, where $r_{max}=\max_{i}\{r_{i}\}$.
The algorithm is so efficient that it runs in time $O(n\log n)$,
where $n=\vert\Gamma\vert$ is the number of the sensors. The second
is generally an linear programming (LP) rounding based approach, which
first transfers MMSM to the fractional cardinality matching problem
and then solves the LP relaxation we propose for the latter problem.
The algorithm approximately solves the decision MMSM problem according
to a solution to the LP relaxation. Similar to the case for the first
algorithm, we show that the algorithm always outputs ``feasible''
if $D\geq D^{*}$. Further, for any instance our algorithm returns
``feasible'', we give a method to construct a real solution for
MMSM, with a maximum relocation distance $D+r_{max}$, by rounding
up a fractional optimum solution to the LP relaxation. The algorithm
has a runtime $O(n^{7}L)$, which is exactly the time of solving the
proposed LP relaxation by Karmakar's algorithm \cite{karmarkar1984new},
where $L$ is the length of the input. As a by-product, we give the
third algorithm for decision MMSM with a maximum relocation distance
$d(OPT)+r_{max}$, and time $O(R^{2}\sqrt{\frac{M}{\log R}})$, where
$R=\sum_{i=1}^{n}2r_{i}$ is sum of the radii of the sensors.

Based on the three above algorithms for the decision problem, the
paper proposes an unified algorithm framework to actually calculate
a solution to MMSM without a given $D$. The time complexity and the
maximum relocation distance are respectively $O(n\log n\log d_{\max})$
and $D^{*}+2r_{max}$ if employing the greedy algorithm; $O(n^{7}L\log d_{max})$
and $D^{*}+r_{max}$ if employing the LP based algorithm, where $d_{max}$
is the maximum distance between the sensors and the barriers. The
runtime $O(n^{7}L\log d_{max})$ can be improved to $O(R^{2}\sqrt{\frac{M}{\log R}}\log d_{max})$
if $M$ is not large, by using the third algorithm based on matching.

Note that, although our algorithm could compute a near-optimal solution
when $D^{*}\gg r_{max}$, the performance guarantee is not as good
when $D^{*}<r_{max}$. So in addition we give a simple factor-2 approximation
algorithm for MMSM, by extending the optimal algorithm for 1D-MMSM
as in paper \cite{chen2013algorithms}. Consequently, the ratio of
our first three algorithms can be improved for the case $D^{*}<r_{max}$,
by combining the factor-2 approximation.

\subsection{Organization of the Paper}

The remainder of the paper is organized as follows: For decision MMSM,
Section 2 gives a greedy algorithm as well as the ratio proof; Section
3 gives an approximation algorithm with an improved maximum relocation
distance $d(OPT)+r_{max}$ using LP rounding technique; Section 4
gives another approximation algorithm with the same maximum relocation
distance guarantee but a different runtime, by using maximum cardinality
matching; Section 5 present the algorithm which actually solve MMSM,
using the algorithm given in Section 2, 3 and 4; Section 6 extends
previous results and develops a factor-2 approximation algorithm with
provable performance guarantee; Section 7 concludes the paper.

\section{A Simple Greedy Algorithm For Decision MMSM}

This section presents an approximation algorithm for any instance
of decision MMSM wrt a given $D$: if the algorithm returns ``infeasible'',
then the instance is truly \emph{infeasible} with respect to $D$;
Otherwise, the instance of MMSM is feasible under the maximum movement
of $D+2r_{max}$, where $r_{max}=\max_{i}\{r_{i}\}$. To show the
performance guarantee of the algorithm, we propose a sufficient condition
for the feasibility of decision MMSM against given $D$.

\subsection{An Approximation Algorithm}

Let $[l_{i},\,g_{i}]$ be the possible coverage range for sensor $i$,
where $l_{i}$ and $g_{i}$ are respectively the leftmost and the
rightmost points of the barrier, i.e. the leftmost and the rightmost
points sensor $i$ can cover by relocating within distance $D$. The
key idea of our algorithm is to cover the barrier from left to right,
using the sensor with minimum $g_{i}$ within the set of sensors which
can cover the leftmost uncovered point with a maximum relocation distance
$D+2r_{max}$.

More detailed, the algorithm is first to compute for each sensor $i$
its possible coverage range $[l_{i},\,g_{i}]$. Let $s$ be the leftmost
point of the uncovered part of the line barrier. Then among the set
of sensors $\{i\vert l_{i}-2r_{max}\leq s\leq g_{i}\}$, the algorithm
repeats selecting the sensor with minimum $g_{i}$ to cover an uncovered
segment of the line barrier starting at $s$. Note that $\{i\vert l_{i}-2r_{max}\leq s\leq g_{i}\}$
is exactly the set of sensors, which are with $g_{i}\geq s$ and can
monitor an uncovered segment starting at $s$ by relocating at most
$D+2r_{max}$ distance. If there is a tie on $g_{i}$, then randomly
pick a sensor within the tie. The selection terminates once the line
barrier is completely covered, or the instance is found infeasible,
i.e. there exists no such $i$ with $l_{i}-2r_{max}\leq s\leq g_{i}$
while the coverage is not done. The algorithm is formally as in Algorithm
\ref{alg:mainA-simple-greedy}.

\begin{algorithm}
\textbf{Input: }A movement distance bound $D\in\mathbb{Z}^{+}$, a
set of sensors $\Gamma=\{1,\,\dots,\,n\}$ with $\{r_{i}\vert i\in[n]^{+}\}$
and $\{(x_{i},\,y_{i})\vert i\in[n]^{+}\}$, in which $r_{i}$ and
$(x_{i},\,y_{i})$ are respectively the sensing radii and the original
position of sensor $i$;

\textbf{Output: }New positions $\{(x_{i}',\,y_{i}')\vert i\in[n]^{+}\}$
for the sensors.

\enskip{}1: Set ${\cal I}:=\Gamma$, $s:=0$; /{*}$s$ is the leftmost
point of the uncovered part of the barrier.{*}/

\enskip{}2: \textbf{For} each sensor $i$ \textbf{do}

\enskip{}3: \quad{}Compute the leftmost position $l_{i}$ and the
rightmost position $g_{i}$, both of which sensor $i$ can monitor;

\enskip{}4: \textbf{While} ${\cal I}\neq\emptyset$ \textbf{do}

\enskip{}5: \quad{}\textbf{If} there exists $i'\in{\cal I}$, such
that $l_{i'}-2r_{max}\leq s\leq g_{i'}$ \textbf{then}

\enskip{}6: \quad{}\quad{}Select $i\in{\cal I}$ for which $g_{i}=\min_{i':\,l_{i'}-2r_{max}\leq s\leq g_{i'}}\{g_{i'}\}$;

\quad{}\quad{}\quad{}\quad{}/{*} Select the sensor with minimum
$g_{i}$ among the sensors $\{i'\vert l_{i'}-2r_{max}\leq s\}$; {*}/

\enskip{}7: \quad{}\quad{}Set $s:=\min\{s+2r_{i},\,g_{i}\}$, ${\cal I}:={\cal I}\setminus\{i\}$,
$x_{i}':=s-r_{i}$, $y_{i}':=0$;

\enskip{}8: \quad{}\textbf{Else}

\enskip{}9: \quad{}\quad{}Return ``infeasible''.

10: Return ``feasible'' the new positions $\{(x'_{i},\,0)\vert i\in[n]^{+}\}$.

\caption{\label{alg:mainA-simple-greedy}A simple greedy algorithm for decision
MMSM.}
\end{algorithm}

For briefness, we will simply say an instance (or the input) of MMSM
instead for the input of Algorithm \ref{alg:mainA-simple-greedy}
in the following paragraphs. Note that Steps 2-3 take time $O(n)$
to compute $l_{i}$ and $g_{i}$ for all the sensors, Steps 4-7 take
$O(n\log n)$ time to assign the sensors to cover the line barrier.
Therefore, we have the time complexity of the algorithm:
\begin{lem}
Algorithm \ref{alg:mainA-simple-greedy} runs in time $O(n\log n)$.
\end{lem}

\subsection{The Ratio of Algorithm \ref{alg:mainA-simple-greedy}}

\begin{figure*}
\includegraphics{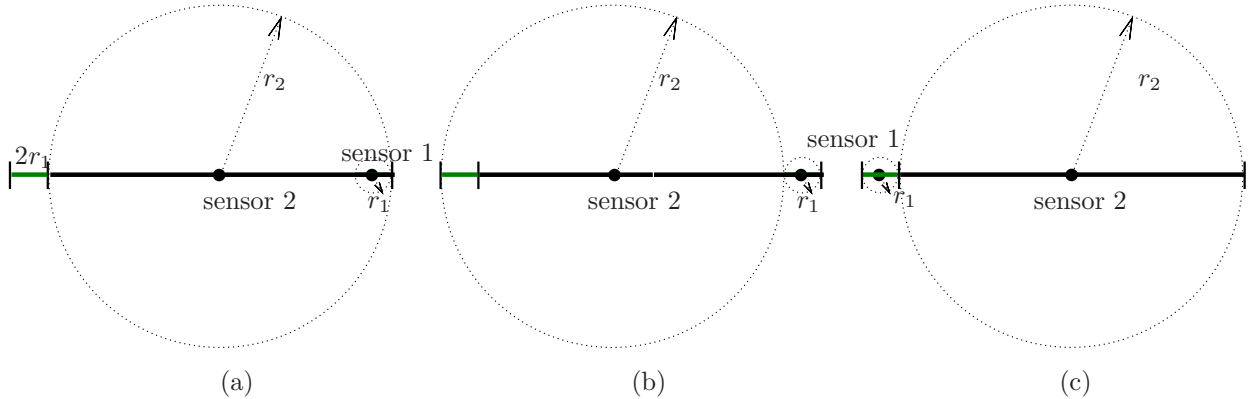}

\caption{\label{fig:A-tight-maingreedalgorithm}A near-tight example for Algorithm
\ref{alg:mainA-simple-greedy}: (a) An instance of decision MMSM:
line barrier is between points $(0,\,0)$ and $(2r_{1}+2r_{2},\,0)$,
while the positions of sensor 1 and 2 are respectively $(2r_{2}+r_{1},\,0)$
and $(2r_{1}+r_{2},\,0)$; (b) The optimal solution with maximum relocation
distance $2r_{1}$; (c) The solution output by Algorithm \ref{alg:mainA-simple-greedy}
with maximum relocation distance $2r_{2}$.}

\end{figure*}

The performance guarantee of Algorithm \ref{alg:mainA-simple-greedy}
is as below:
\begin{thm}
\label{thm:greedyfinalthr}Let $D^{*}$ be the distance of an optimal
solution. If $D\geq D^{*}$, then Algorithm \ref{alg:mainA-simple-greedy}
will return a solution with maximum relocation distance $D+2r_{max}$,
where $r_{max}=\max_{i}\{r_{i}\}$.
\end{thm}
According to Algorithm \ref{alg:mainA-simple-greedy}, we never move
a sensor out of the range $[l_{i}-2r_{max},\,g_{i}]$. It remains
to show Algorithm \ref{alg:mainA-simple-greedy} will always return
a feasible solution when $D\geq D^{*}$. For this goal, we will give
a sufficient condition for the feasibility of decision MMSM. Below
are two notations needed for the tasks:
\[
\lambda(i,\,D,\,x,\,x')=\begin{cases}
0 & g_{i}\leq x\,\mbox{or}\,l_{i}\geq x'\\
\min\{2r_{i},\,\min\{x',\,g_{i}\}-\max\{x,\,l_{i}\}\} & Otherwise
\end{cases}
\]

and

\[
\sigma(i,\,D,\,{\cal S})=\begin{cases}
0 & g_{i}\leq x\,\mbox{OR}\,l_{i}\geq x'\\
\min\{2r_{i},\,\sum_{[x,x']\in{\cal S}}\lambda(i,\,D^{*},\,x,\,x')\} & Otherwise
\end{cases}
\]
where ${\cal S}$ is a set of \emph{disjoint} segments of the line
barrier. Intuitionally $\lambda(i,\,D,\,x,\,x')$ is the maximum coverage
which sensor $i$ can provide for segment $[x,\,x']$, and $\sigma(i,\,D,\,{\cal S})$
is the sum of the coverage that sensor $i$ can provide for all the
segments in ${\cal S}$. Then a simple necessary condition for the
feasibility of an instance of MMSM of as below:
\begin{prop}
\label{prop:necessary}If an instance of decision MMSM is feasible
wrt $D$, then $\sum_{i}\sigma(i,\,D,\,{\cal S})\geq\sum_{[x_{j},\,x_{j}']\in{\cal S}}\vert x_{j}'-x_{j}\vert$
must hold for any set of disjoint segments ${\cal S}=\{[x_{i},\,x_{i}']\vert0\leq i\leq M\}$.
\end{prop}
Intuitionally, the above proposition states that the sum of the sensor
coverage length must be not less than the length of the barrier segments
to cover. The correctness of the above lemma is obviously, since a
feasible relocation assignment must satisfy the condition. However,
this is not a sufficient condition for the feasibility of decision
MMSM (A counter example is as depicted in Figure \ref{fig:A-tight-exampleforLP}
(a): For $D=r_{1}$, the necessary condition holds for the given instance
while the instance is actually infeasible). However, if $2r_{max}$
more relocation distance is allowed as in Algorithm \ref{alg:mainA-simple-greedy},
we have the following lemma:
\begin{lem}
\label{lem:sufficientcoreforgreedy}If $\sum_{i}\sigma(i,\,D,\,{\cal S})\geq\sum_{[x_{j},\,x_{j}']\in{\cal S}}\vert x_{j}'-x_{j}\vert$
holds for every disjoint segments set ${\cal S}$, then the instance
of decision MMSM is feasible under maximum relocation distance $D+2r_{max}$,
$r_{max}=\max_{i}\{r_{i}\}$.
\end{lem}
\begin{IEEEproof}
We need only to show that if $\sum_{i}\sigma(i,\,D,\,{\cal S})\geq\sum_{[x_{j},\,x_{j}']\in{\cal S}}\vert x_{j}'-x_{j}\vert$
holds at the beginning of Algorithm \ref{alg:mainA-simple-greedy},
then $\sum_{i}\sigma(i,\,D,\,{\cal S})\geq\sum_{[x_{j},\,x_{j}']\in{\cal S}}\vert x_{j}'-x_{j}\vert$
remains true in each step of Algorithm \ref{alg:mainA-simple-greedy}.

Suppose the lemma is not true. Let the step of picking sensor $i$
be the first $\sum_{i}\sigma(i,\,D,\,{\cal S})\geq\sum_{[x_{j},\,x_{j}']\in{\cal S}}\vert x_{j}'-x_{j}\vert$
becomes \emph{false}. Then there must exist $x$ and $x'$, such that
$\sum_{j\in{\cal I}}\sigma(j,\,D,\,x,\,x')\geq\vert x'-x\vert$ and
$\sum_{j\in{\cal I}\setminus i}\sigma(j,\,D,\,x,\,x')<\vert x'-x\vert$.
We analysis all cases wrt all the possible orders of $g_{i}$, $s+2r_{i}$
and $x$ in the line barrier, and show that contradictions exist in
every case.

\begin{enumerate}
\item $g_{i}\leq x$:

In this case, $\sigma(i,\,D,\,[x,\,x'])=0$ holds, i.e. sensor $i$
does not cover any portion of $[x,\,x']$. So we have
\[
\sum_{j\in{\cal I}\setminus\{i\}}\sigma(j,\,D,\,x,\,x')=\sum_{j\in{\cal I}}\sigma(j,\,D,\,[x,\,x'])\geq\vert x'-x\vert,
\]
which contradicts with $\sum_{j\in{\cal I}\setminus\{i\}}\sigma(j,\,D,\,[x,\,x'])<\vert x'-x\vert$.
\item $x<g_{i}\leq s+2r_{i}$:

In this case, sensor $i$ will actually cover $[s,\,s+2r_{i}]$, and
hence $\sigma(i,\,D,\,[x,\,x'])=\min\{g_{i},\,x'\}-x$ is the actual
coverage that sensor $i$ can contribute to $\sum_{j\in{\cal I}}\sigma(j,\,D,\,[x,\,x'])$.
So
\begin{equation}
\sum_{j\in{\cal I}\setminus\{i\}}\sigma(j,\,D,\,[x,\,x'])=\sum_{j\in{\cal I}}\sigma(j,\,D,\,[x,\,x'])-(\min\{g_{i},\,x'\}-x).\label{eq:eq1}
\end{equation}
Then by combining $\sum_{j\in{\cal I}}\sigma(j,\,D,\,x,\,x')\geq x'-x$
with Inequality (\ref{eq:eq1}), we have

\begin{equation}
\sum_{j\in{\cal I}\setminus\{i\}}\sigma(j,\,D,\,[x,\,x'])\geq x'-x-(\min\{g_{i},\,x'\}-x)=x'-\min\{g_{i},\,x'\}\label{eq:eq3}
\end{equation}

On the other hand, the length of the portion of $[x,\,x']$ that needs
to be covered is apparently $\vert x'-x\vert-(\min\{g_{i},\,x'\}-x)$.
From the infeasibility of the remaining sensors in ${\cal I}$, we
have
\begin{equation}
\sum_{j\in{\cal I}\setminus\{i\}}\sigma(j,\,D,\,[x,\,x'])<x'-x-(\min\{g_{i},\,x'\}-x)=x'-\min\{g_{i},\,x'\}.\label{eq:reve}
\end{equation}
 A contradiction arises by comparing Inequality (\ref{eq:eq3}) and
(\ref{eq:reve}).
\item $g_{i}>x$ and $g_{i}>s+2r_{i}$:

Assume that $s+2r_{i}\leq x$. This assumption is without loss of
generality, since otherwise from $\sum_{j\in{\cal I}\setminus\{i\}}\sigma(j,\,D,\,x,\,x')<\vert x'-x\vert$
and the fact that $[x,\,s+2r_{i}]$ is already covered by sensor $i$,
we have $\sum_{j\in{\cal I}\setminus\{i\}}\sigma(j,\,D,\,s+2r_{i},\,x')<x'-(s+2r_{i})$.
That is, we need only to set $x=s+2r_{i}$, and obtain contractions
similar as this case. We will show that $\sum_{j\in{\cal I}\setminus\{i\}}\sigma(j,\,D,\,s+2r_{i},\,x')\geq x'-(s+2r_{i})$
actually holds and get a construction

By inductions, we have

\[
\sum_{j\in{\cal I}}\sigma(j,\,D,\,\{[s,\,s+2r_{i}],\,[x,\,x']\})\geq2r_{i}+x'-x.
\]

That is,

\[
\sum_{j\in{\cal I}}\sigma(j,\,D,\,\{[s,\,s+2r_{i}],\,[x,\,x']\})-2r_{i}\geq x'-x.
\]

So

\begin{equation}
\sum_{j\in{\cal I}\setminus\{i\}}\sigma(j,\,D,\,\{[s,\,s+2r_{i}],\,[x,\,x']\})\geq x'-x.\label{eq:suff}
\end{equation}

Assume that sensor $j\neq i$ is a sensor which can contribute to
both $\sigma(j,\,D,\,[s,\,s+2r_{i}])$ and $\sigma(j,\,D,\,[x,\,x'])$.
Let $\phi(j,\,D,\,[s,\,s+2r_{i}])$ and $\phi(j,\,D,\,[x,\,g_{i}])$
be the portion that sensor $j$ actually contributes $[s,\,s+2r_{i}]$
and $[x,\,g_{i}]$, respectively, within $\sum_{j\in{\cal I}}\sigma(j,\,D,\,([s,\,s+2r_{i}],\,[x,\,x']))$.
We need only to show that $\sum_{j\in{\cal I}\setminus\{i\}}\phi(j,\,D,\,[s,\,s+2r_{i}])$
is sufficient to be relocated to compensate all the coverage sensor
$i$ contributes to $[x,\,g_{i}]$.

Since the chosen sensor $i$ is with smallest $g_{i}$ within all
the sensors of $l_{i}-2r_{max}\leq s$, sensor $j\in{\cal I}\setminus\{i\}$
is with $g_{j}\geq g_{i}$. So the potion of sensor $j$ covering
$[s,\,s+2r_{i}]$, i.e. $\phi(j,\,D,\,[s,\,s+2r_{i}])$, can all be
relocated to cover any portion of $[x,\,g_{i}]$. So $\phi(j,\,D,\,[s,\,s+2r_{i}])+\phi(j,\,D,\,[x,\,\min\{g_{i},\,x'\}])$
is actually the portion of the cover that sensor $j$ can contributes
to $[x,\,g_{i}]$. Therefore, we have
\begin{eqnarray}
\sum_{j\in{\cal I}\setminus\{i\}}\sigma(j,\,D,\,[x,\,x']) & = & \sum_{j\in{\cal I}\setminus\{i\}}(\phi(j,\,D,\,[s,\,s+2r_{i}])\label{eq:sum}\\
+\phi(j,\,D,\,[x,\,\min\{g_{i},\,x'\}]) & + & \sum_{j\in{\cal I}\setminus\{i\}}\phi(j,\,D,\,[\min\{g_{i},\,x'\},\,x']\}).\nonumber
\end{eqnarray}
On the other hand, we have

\begin{eqnarray}
 & \sum_{j\in{\cal I}\setminus\{i\}}\sigma(j,\,D,\,\{[s,\,s+2r_{i}],\,[x,\,x']\}) & \leq\sum_{j\in{\cal I}\setminus\{i\}}\phi(j,\,D,\,[s,\,s+2r_{i}])\label{eq:theother}\\
+ & \sum_{j\in{\cal I}\setminus\{i\}}\phi(j,\,D,\,[x,\,\min\{g_{i},\,x'\}]) & +\sum_{j\in{\cal I}\setminus\{i\}}\phi(j,\,D,\,[\min\{g_{i},\,x'\},\,x']\}).\nonumber
\end{eqnarray}

Combining Inequality (\ref{eq:suff}), (\ref{eq:sum}) and (\ref{eq:theother}),
we have $\sum_{j\in{\cal I}\setminus\{i\}}\sigma(j,\,D,\,[x,\,x'])\geq\sum_{j\in{\cal I}\setminus\{i\}}\sigma(j,\,D,\,\{[s,\,s+2r_{i}],\,[x,\,x']\})\geq x'-x$,
which contradicts with the assumption.
\end{enumerate}
\end{IEEEproof}
Now we will prove Theorem \ref{thm:greedyfinalthr}. If $D\geq D^{*}$,
then the decision MMSM is feasible, and hence following Proposition
\ref{prop:necessary} $\sum_{i}\sigma(i,\,D,\,{\cal S})\geq\sum_{[x_{j},\,x_{j}']\in{\cal S}}\vert x_{j}'-x_{j}\vert$
holds for every ${\cal S}$ at the beginning of Algorithm \ref{alg:mainA-simple-greedy}.
Then from Lemma \ref{lem:sufficientcoreforgreedy}, we immediately
have the instance is feasible under relocation distance bound $D$,
which completes the proof of Theorem \ref{thm:greedyfinalthr}.
\begin{cor}
The performance guarantee of Algorithm \ref{alg:mainA-simple-greedy}
in Theorem \ref{thm:greedyfinalthr} is nearly tight.
\end{cor}
As the example depicted in Figure \ref{fig:A-tight-maingreedalgorithm},
$d(OPT)=2r_{1}$ while the output of the algorithm is with a maximum
relocation distance $2r_{2}$. So when $r_{2}\gg r_{1}$, $d(OPT)+2r_{max}\thickapprox2r_{1}+2r_{2}$,
and hence the analysis of Algorithm \ref{alg:mainA-simple-greedy}
is nearly tight in Theorem \ref{thm:greedyfinalthr}.

\section{An LP-based Approximation for Decision MMSM}

This section will give an LP-based approximation algorithm to determine
whether an instance of the decision MMSM problem is feasible. The
algorithm first transfers the instance to a corresponding instance
of decision DMMSM, and then an instance of the fractional cardinality
matching problem with a proposed LP relaxation. Our algorithm answers
``feasible'' or ``infeasible'' according to the computed optimum
solution of the LP relaxation. We show that if our algorithm returns
\emph{feasible}, then a solution to MMSM can be constructed under
the maximum relocation distance $D+r_{max}$ by rounding up a fractional
optimum solution to the relaxation.

\subsection{Transferring to an Instance of DMMSM }

The key idea of the transfer is first to compute $l_{i}$ and $g_{i}$
for each $i$ wrt to the given $D$, and then add $(l_{i},\,0)$ and
$(g_{i},\,0)$ to the barrier as two vertices on the line. That is,
$V=\{(l_{i},\,0)\vert i\in[n]^{+}\}\cup\{(g_{i},\,0)\vert i\in[n]^{+}\}$.
W.l.o.g. assume that the vertices of $V$ appear on the line barrier
in the order of $v_{0},\,\dots,\,v_{|V|}$, from left to right on
the line barrier. Then the algorithm adds an edge between every pair
of $v_{i}$ and $v_{i+1}$. So $E=\{e_{i}(v_{i},\,v_{i+1})\vert i\in[\vert V\vert-1]^{+}\}$.
Formally the transfer is as in Algorithm \ref{alg:The-transfering-algorithmfromMMSMtoDMMSM}.

\begin{algorithm}
\textbf{Input:} An instance of MMSM;

\textbf{Output: }$G=(V,\,E)$, an instance of DMMSM.

\enskip{}1: Set $V:=\emptyset$ and $E:=\emptyset$;

\enskip{}2: \textbf{For} each edge $i\in S$ \textbf{do}

\enskip{}3: \quad{}Compute $l_{i}$ and $g_{i}$;

\enskip{}4:\textbf{ }\quad{}$V\leftarrow V\cup\{(l_{i},\,0),\,(g_{i},\,0)\}$;\textbf{ }

\enskip{}5: Number the vertices of $V$, such that the vertices appear
in the barrier from left to right in the order of $v_{0},\,\dots,\,v_{|V|}$;

\enskip{}6: Set $E:=\{e_{j}(v_{j},\,v_{j+1})\vert j\in[\vert V\vert-1]^{+}\}$;

\enskip{}7:\textbf{ }Return $G=(V,\,E)$.

\caption{\label{alg:The-transfering-algorithmfromMMSMtoDMMSM}The transferring
algorithm. }
\end{algorithm}

For the time complexity and the size of the graph, we have:
\begin{lem}
Algorithm \ref{alg:The-transfering-algorithmfromMMSMtoDMMSM} runs
in $O(n\log n)$ time, and output a graph $G$ with $\vert V\vert=O(n)$
and $\vert E\vert=O(n)$.
\end{lem}
According to the algorithm, $\vert V\vert=O(n)$ and $\vert E\vert=O(n)$
hold trivially. Algorithm \ref{alg:The-transfering-algorithmfromMMSMtoDMMSM}
takes in $O(n\log n)$ time to sort (number) the vertices of $V$
in Step 5, since $\vert V\vert=O(n)$. Other steps of the algorithm
takes trivial time compared to the sorting. So the total runtime of
Algorithm \ref{alg:The-transfering-algorithmfromMMSMtoDMMSM} is $O(n\log n)$.
\begin{lem}
An instance of MMSM is feasible under $D$ if and only if its corresponding
DMMSM instance produced by Algorithm \ref{alg:The-transfering-algorithmfromMMSMtoDMMSM}
is feasible under $D$.
\end{lem}
\begin{IEEEproof}
According to Algorithm \ref{alg:The-transfering-algorithmfromMMSMtoDMMSM}
and following the definition of MMSM and DMMSM, a solution to an instance
of MMSM is obviously a solution to the corresponding instance of DMMSM,
and vice versa. So an instance of MMSM is feasible, iff its corresponding
DMMSM instance is feasible.
\end{IEEEproof}

\subsection{Fractional Maximum Cardinality Matching wrt DMMSM}

Let $\vert e_{i}\vert=v_{i+1}-v_{i}$ be the length of edge $e_{i}$.
Assuming that $v_{i_{1}}=l_{i}$ and $v_{i_{2}}=g_{i}$, we set $J_{i}=\{e_{j}\vert j=i_{1},\,i_{1}+1,\,\dots,\,i_{2}-1\}$.
Then the linear programming relaxation (LP1) for DMMSM is as below:

\[
\max\sum_{i\in[n]^{+}}\sum_{j\in J_{i}}x_{i,\,j}
\]

subject to

\begin{eqnarray}
\sum_{j\in J_{i}}x_{i,j} & \leq2r_{i} & \forall i\in\{1,\,\dots,\,n\}\label{eq:sensoratmostone}\\
\sum_{i\in[n]^{+}}x_{i,\,j} & \leq\vert e_{j}\vert & \forall j\in\{1,\,\dots,\,\vert V\vert\}\label{eq:coverrequirement}\\
0\leq x_{i,\,j} & \leq\vert e_{j}\vert & \forall i\in\{1,\,\dots,\,n\},\,j\in J_{i}\label{eq:capacitybd}
\end{eqnarray}
where $\sum_{j\in J_{i}}x_{i,j}\leq2r_{i}$ is because a sensor $i$
can at most cover length $2r_{i}$ of the barrier, and Inequality
(\ref{eq:coverrequirement}) is because the covered length of each
edge $e_{i}$ (segment) is at most $\vert e_{j}\vert$.

Our algorithm determines decision DMMSM according to the computed
optimum solution, say $\mathbf{x}$, to LP1: the algorithm outputs
``feasible'' if $\sum_{i\in[n]^{+}}\sum_{j\in J_{i}}x_{i,\,j}=M$,
and outputs ``infeasible'' otherwise.

\begin{algorithm}
\textbf{Input:} An instance of DMMSM;

\textbf{Output: }Answer whether the instance is feasible.

\enskip{}1: Solve LP1 against the instance of DMMSM by Karmakar's
algorithm as in \cite{karmarkar1984new}, and obtain an optimal solution
$\mathbf{x}$;

\enskip{}2: \textbf{If} $\sum_{i\in[n]^{+}}\sum_{j\in J_{i}}x_{i,\,j}=M$
according to $\mathbf{x}$ \textbf{then}

\enskip{}3: \quad{}Return ``feasible'';

\enskip{}4:\textbf{ else}

\enskip{}5: \quad{}Return ``infeasible'';

\caption{\label{alg:mainLProundDMMSM}The determining algorithm for decision
DMMSM. }
\end{algorithm}

It is known that there exist polynomial-time algorithms for solving
linear programs. In particular, using Karmakar's algorithm to solve
LP1 will take $O(n^{7}L)$ time \cite{karmarkar1984new}, since there
are $O(n^{2})$ variables totally in LP1.
\begin{lem}
Algorithm \ref{alg:mainLProundDMMSM} runs in time $O\left(n^{7}L\right)$.
\end{lem}
It is worth to note that the simplex algorithm has a much better practical
performance than Karmakar's algorithm \cite{korte2002combinatorial}.
So using the simplex method the algorithm would be faster than $O(n^{7}L)$
in real world applications.

\begin{figure*}
\includegraphics[width=0.85\textwidth]{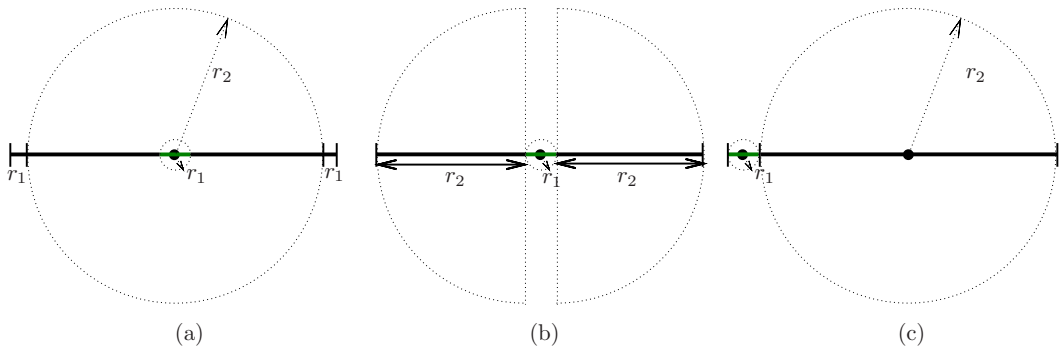}

\caption{\label{fig:A-tight-exampleforLP}The integrality gap: (a) An instance
of decision MMSM: line barrier is between points $(0,\,0)$ and $(2r_{1}+2r_{2},\,0)$,
while the positions of sensor 1 and 2 are both $(r_{1}+r_{2},\,0)$;
(b) Algorithm \ref{alg:mainLProundDMMSM} outputs ``feasible'' if
$D\geq r_{2}$; (c) The optimal maximum relocation distance is actually
$r_{1}$.}
\end{figure*}

The performance guarantee of Algorithm \ref{alg:mainLProundDMMSM}
is as given in the following theorem, whose proof will be given in
next subsection.
\begin{thm}
\label{thm:LPbasedratio}If Algorithm \ref{alg:mainLProundDMMSM}
returns ``infeasible'', then the instance of DMMSM is truly infeasible
under the given $D$; Otherwise, the instance of DMMSM is truly feasible
under the maximum relocation distance $D+r_{max}$, where $r_{max}=\max_{i}\{r_{i}\}$.
\end{thm}
\begin{cor}
The performance guarantee given in Theorem \ref{thm:LPbasedratio}
is nearly tight for Algorithm \ref{alg:mainLProundDMMSM}.
\end{cor}
From Figure \ref{fig:A-tight-exampleforLP}, an optimal fractional
solution to LP1 is with a maximum relocation distance $r_{1}$, while
a true optimal solution for MMSM must be with a maximum relocation
distance $r_{2}$. Thus, the integral gap for LP1 is $r_{2}-r_{1}\approx r_{max}$
when $r_{2}\gg r_{1}$. That is, for any fixed $\epsilon>0$, the
maximum relocation distance increment could be larger than $r_{max}-\epsilon$
for rounding an optimum solution of LP1 to a true solution of MMSM.
Therefore, the ratio is nearly-tight for Algorithm \ref{alg:mainLProundDMMSM}.

\subsection{Proof of Theorem \ref{thm:LPbasedratio} }

This subsection will prove Theorem \ref{thm:LPbasedratio} by showing
a fractional optimum solution to LP1 can be rounded up to an integral
solution of DMMSM with a maximum relocation distance $D+r_{max}$.

Let $\mathbf{x}$ be an optimum solution to LP1. Recall that $x_{ij}$
is the (fractional) portion sensor $i$ covering edge $j$, the key
idea of our algorithm is to aggregate the portions of sensor $i$
covering different $j$s, such that the portions combine to a line
segment which $i$ can cover within movement $D+r_{max}$.

Our algorithm is composed by two parts. The first part is called \emph{pre-aggregation}
which rounds $x_{ij}$ to $1$ in a ``pseudo'' way. More precisely,
assume that $0<x_{i_{1},\,j},\,\dots,\,x_{i_{h},\,j}<0$ are the variables
shares edge $e_{j}$. Then the \emph{pre-aggregation} divides $e_{j}$
to a set of sub-edges $\{e_{j,\,i_{1}},\,\dots,e_{j,\,i_{h}}\}$,
in which $\vert e_{j,\,i_{l}}\vert=\vert e_{j}\vert\cdot x_{i_{l},\,j}$.
We set $x_{i_{l},\,j,\,l}=1$, which is, edge $e_{j,\,i_{l}}$ completely
covered by sensor $i_{l}$.

\begin{figure*}
\includegraphics{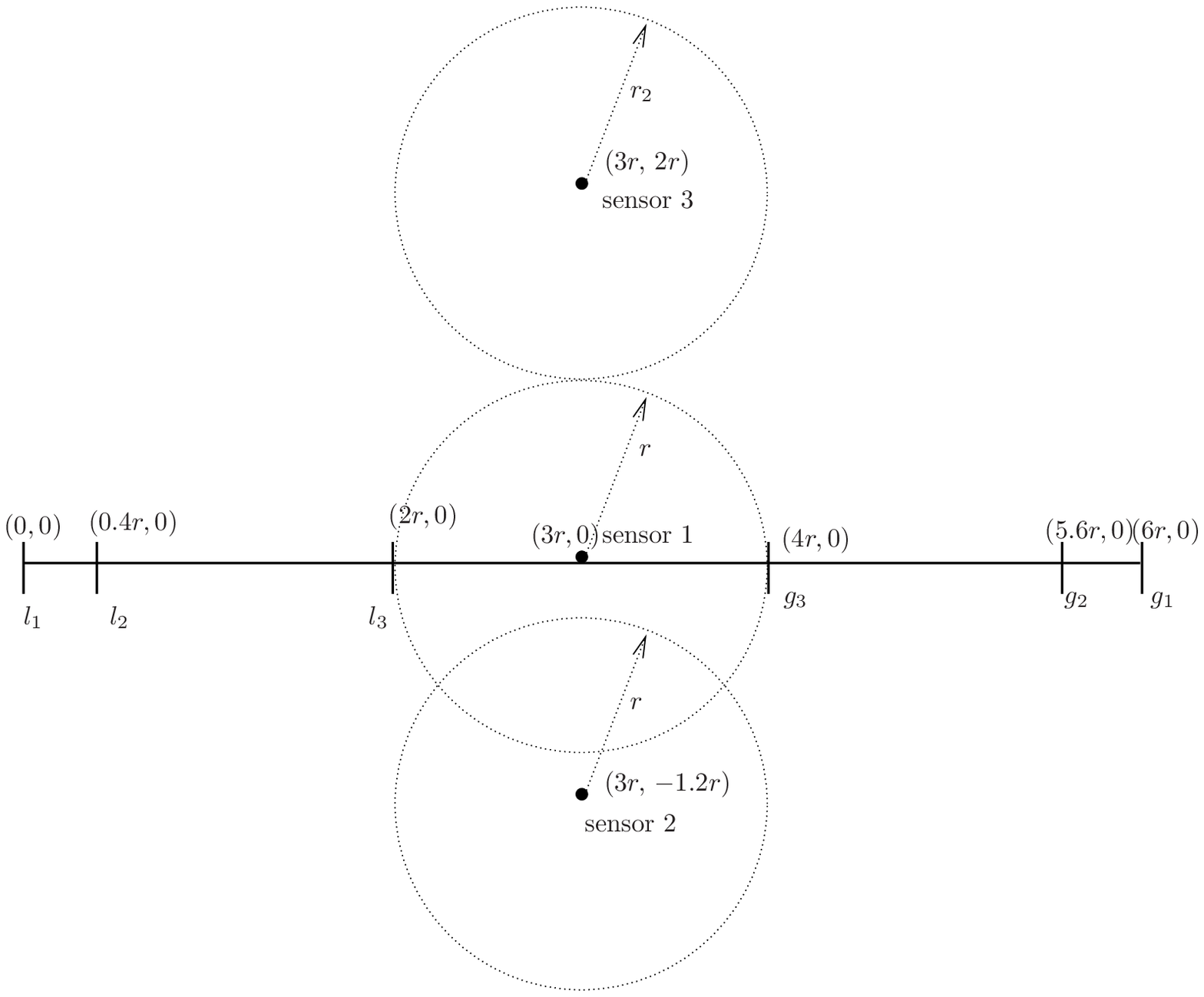}

\caption{\label{fig:ori}An instance of decision MMSM: Sensors 1, 2 and 3 are
with identical sensing radii $r$, and positions $(3r,\,0)$, $(3r,\,-1.2r)$,
$(3r,\,2r)$, respectively. The barrier to cover is $[0,\,6r]$ and
the given $D=2r$.}
\end{figure*}

\begin{figure*}
\includegraphics{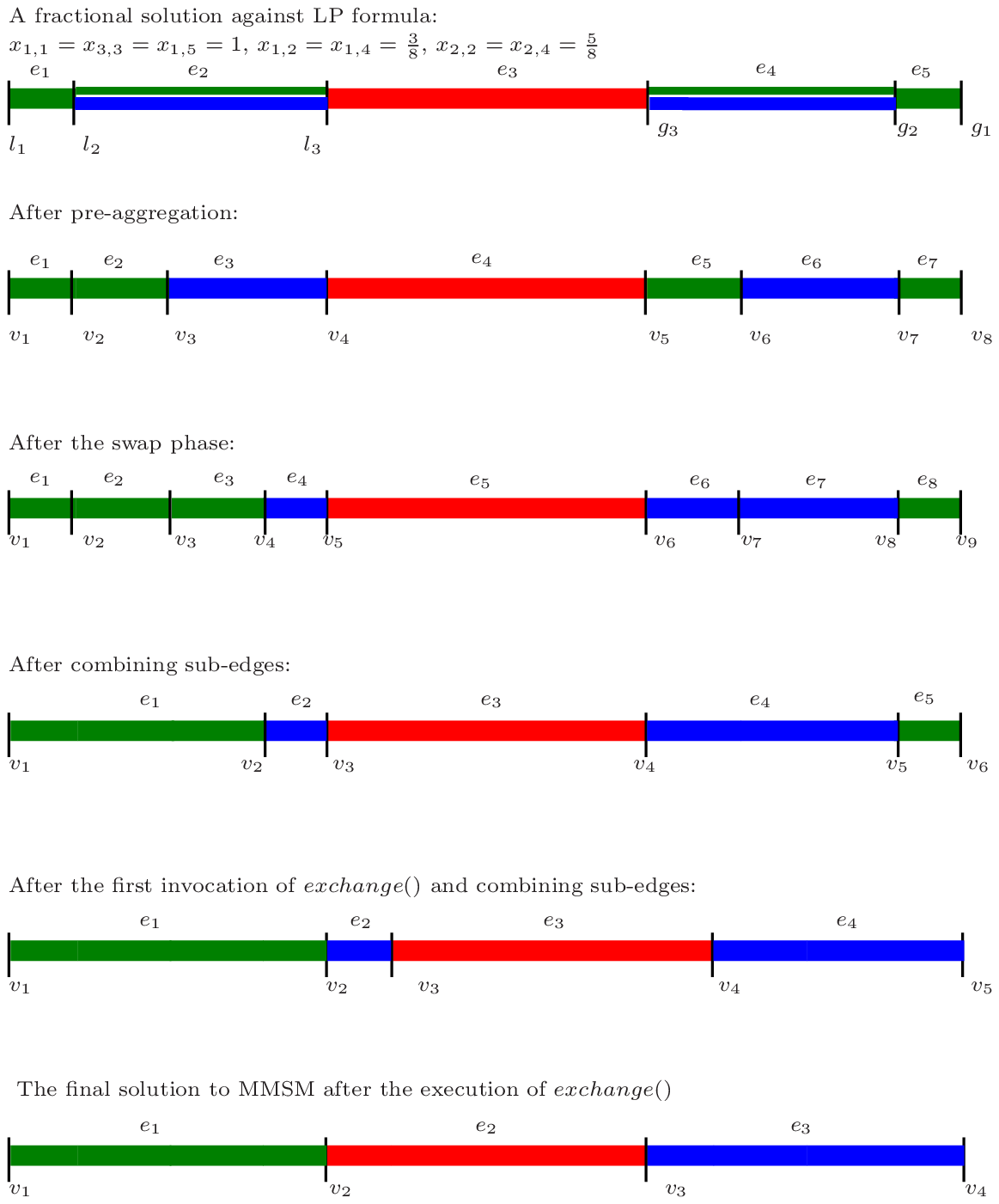}

\caption{\label{fig:exe} Execution of Algorithm \ref{alg:rounding} against
the instance as in Figure \ref{fig:ori}.}
\end{figure*}

Let ${\cal S}_{i}$ be the set of sub-edges covered by sensor $i$
accordingly. The second part, which is called \emph{aggregation,}
aggregates the edges of ${\cal S}_{i}$ for each $i$ such that the
edges covered by an identical sensor will connect together. The \emph{aggregation}
starts from the following simple observation whose correctness is
obviously:
\begin{prop}
Let $i$ and $i'$ be two sensors. Let ${\cal S}_{i}$ and ${\cal S}_{i}'$
be the set of edges covered by sensor $i$ and $i'$, respectively.
Then, for any sub-edges $j_{1},\,j_{2}\in{\cal S}_{i}$ and $j_{1}',\,j_{2}'\in{\cal S}_{i'}$,
w.l.o.g. assuming $j_{1}<j_{1}'$, $j_{1}<j_{2}$ and $j_{1}'<j_{2}'$,
then exactly one of the following cases holds: (1) $j_{1}<j_{2}<j_{1}'<j_{2}'$
; (2) $j_{1}<j_{1}'<j_{2}<j_{2}'$ ; (3) $j_{1}<j_{1}'<j_{2}'<j_{2}$.
\end{prop}
The key observation of the \emph{aggregation} is that if case (2)
and (3) in the above proposition can be eliminated, then the set of
edges in ${\cal S}_{i}$ are aggregated together that they can be
truly monitored by sensor $i$. So the \emph{aggregation} is accordingly
composed by two phases called the \emph{swap phase} and the \emph{exchange}
step, which are to eliminate case (2) and (3), respectively. Formally,
the algorithm is as in Algorithm \ref{alg:rounding} (An example of
such rounding depicted as in Figure \ref{fig:ori} and \ref{fig:exe}).

\begin{algorithm}
Input: $\mathbf{x}$, an optimum solution to LP1 with $\sum_{i\in[n]^{+}}\sum_{j\in J_{i}}x_{i,\,j}=M$;

Output: Relocate positions for the sensors.

\enskip{}1: Run pre-aggregation: compute ${\cal S}_{i}$ for each
sensor $i$ according to $\mathbf{x}$;

\enskip{}\quad{} /{*}${\cal S}_{i}$ contains the sub-edges covered
by sensor $i$. {*}/

\enskip{}2: Set ${\cal S}=\{{\cal S}_{i}\vert i\in[n]^{+}\}$;

\enskip{}3: Swap(${\cal S}$); /{*}The \emph{swap phase}.{*}/

\enskip{}4: Combine every pairs of adjacent edges for each ${\cal S}_{i}\in{\cal S}$;

\enskip{}5: Exchange (${\cal S}$); /{*}The \emph{exchange} step.
{*}/

\enskip{}6: \textbf{For} $i=1$ to $n$ \textbf{do}

\enskip{}\quad{}\quad{} Compute $(x_{i}',\,0)$ such that every
sub-edge of ${\cal S}_{i}$ is with the range $[x_{i}'-r_{i},\,x_{i}'+r_{i}]$
while $\sqrt{(x_{i}'-x_{i})^{2}+y_{i}^{2}}$ attains minimum;

\enskip{}7: Return $\{(x'_{i},\,0)\vert i=1,\,\dots,\,n\}$.

\caption{\label{alg:rounding}An approximation algorithm for rounding an optimum
solution to LP1.}
\end{algorithm}

Note that if two adjacent sub-edges belong to the same ${\cal S}_{i}$,
say $e_{j_{1}},\,e_{j_{2}}\in{\cal S}_{i}$ with $v_{j_{1}+1}=v_{j_{2}}$,
then we can combine the two sub-edges as one, since they are segments
both covered by sensor $i$. So Step 4 of Algorithm \ref{alg:rounding}
is actually to set ${\cal S}_{i}:={\cal S}_{i}\setminus\{e_{j_{1}},\,e_{j_{2}}\}\cup e(v_{j_{1}},\,v_{j_{2}+1})$
for every such pair of adjacent edges of ${\cal S}_{i}$ for every
$i$.

The \emph{swap }phase, as in Step 3 of Algorithm \ref{alg:rounding},
will eliminate case (2) (i.e. $j_{1}<j_{1}'<j_{2}<j_{2}'$) by swapping
the coverage sensors of the edges without causing any increment on
the maximum relocation distance. The observation inspiring the swap
is that for any $j_{1}<j_{1}'<j_{2}<j_{2}'$, we can swap the two
sensors covering $j_{1}'$ and $j_{2}$ without increasing the maximum
relocation distance. More precisely, we cover a portion of $\min\{\vert j_{1}'\vert,\,\vert j_{2}\vert\}$
of $j_{1}'$ with $i$ and cover a portion of $\min\{\vert j_{1}'\vert,\,\vert j_{2}\vert\}$
of $j_{2}$ with $i'$. The formal layout of the \emph{swap phase}
is as in Algorithm \ref{alg:swap}.

\begin{algorithm}
\enskip{}1: \textbf{While} ${\cal S}\neq\emptyset$ \textbf{do}

\enskip{}2: \quad{}Find $S_{i}\in{\cal S}$ that contains the leftmost
edge;

\enskip{}3: \quad{}\textbf{For} $h=1$ to $\vert{\cal S}_{i}\vert-1$\textbf{do}

\enskip{}4: \quad{}\quad{}\textbf{For} ${\cal S}_{i'}\in{\cal S}\setminus{\cal S}$
\textbf{do}

\enskip{}5: \quad{}\quad{}\quad{}Find a pair of sub-edges $e_{u},\,e_{w}\in{\cal S}_{i'}$
such that $j_{h}<u<j_{h+1}<w$ holds;

\enskip{}\enskip{}\quad{}\quad{}\quad{}\quad{}/{*}Recall that
the two endpoints of edge $e_{u}$ is $v_{u}$ and $v_{u+1}$.{*}/

\enskip{}6: \quad{}\quad{}\quad{}\textbf{If} no such $e_{u},\,e_{w}$
exists \textbf{then}

\enskip{}7: \quad{}\quad{}\quad{}\quad{}break;

\enskip{}8: \quad{}\quad{}\quad{}\textbf{If} $\vert e_{u}\vert\geq\vert e_{j_{h+1}}\vert$
\textbf{then}

\enskip{}9: \quad{}\quad{}\quad{}\quad{}Add vertex $p=(x(v_{u})+\vert e_{j_{h+1}}\vert,\,0)$
to $G$; /{*}$x(v_{u})$ is the $x$coordinator of $v_{u}$.{*}/

10: \quad{}\quad{}\quad{}\quad{}${\cal S}_{i}:=S_{i}\setminus\{e_{j_{h+1}}\}\cup\{e(v_{u},\,p)\}$
and ${\cal S}_{i'}:=S_{i'}\setminus\{e_{u}\}\cup\{e_{j_{h+1}}\}\cup e(p,\,v_{u+1})$;

11: \quad{}\quad{}\quad{}\textbf{Else}

12: \quad{}\quad{}\quad{}\quad{}Add vertex $p=(x(v_{j_{h+1}+1})-\vert e_{u}\vert,\,0)$
to $G$;

13: \quad{}\quad{}\quad{}\quad{}${\cal S}_{i'}:=S_{i'}\setminus\{e_{u}\}\cup\{e(p,\,v_{j_{h+1}+1})\}$
and ${\cal S}_{i}:=S_{i}\setminus\{e_{j_{h+1}}\}\cup\{e_{u}\}\cup e(v_{j_{h+1}},\,p)$;

14: \quad{}\quad{}\quad{}Update the numbering of the vertices and
the edges accordingly;

15: \quad{}${\cal S}:={\cal S}\setminus{\cal S}_{i}$;

\caption{\label{alg:swap}Swap(${\cal S}$).}
\end{algorithm}

Note that Steps 8-13 will add new vertices and edges to the graph,
so $\sum_{i=1}^{n}\vert{\cal S}_{i}\vert$ may increases. However,
we can always guarantee $\sum_{i=1}^{n}\vert{\cal S}_{i}\vert\leq n\vert E\vert$.
Since otherwise, following the pigeonhole principle, there must exist
two sub-edges, say $e_{j_{1}}$ and $e_{j_{2}}$ which are in an identical
${\cal S}_{i}$ and within the range of an identical edge of $E$.
Then, such two sub-edges $e_{j_{1}},\,e_{j_{2}}$ can be combined
as one, by setting $x(v_{j_{1}+1})=x(v_{j_{1}+1})+\vert e_{j_{2}}\vert$
and move every sub-edge between $e_{j_{1}}$ and $e_{j_{2}}$ to right
with an offset with length $\vert e_{j_{2}}\vert$. Clearly, following
the meaning of a sensor covering edges as in the definition of DMMSM,
these movement does not cause any increment on the relocation distance
of each sensor.

In Algorithm \ref{alg:swap}, the \emph{while}-loop iterates for $O(n)$
times and the outer \emph{for}-loop iterates for $O(n)$ times. Since
the inner \emph{for}-loop iterates for at most $O(\sum_{i=1}^{n}\vert{\cal S}_{i}\vert)$
times. Then from $\sum_{i=1}^{n}\vert{\cal S}_{i}\vert\leq n\vert E\vert=O(n^{2})$,
the \emph{for-loops iterates at most $O(n^{2})$ time. }So we have
the total runtime of the \emph{swap phase}:
\begin{lem}
\label{lem:swaptime}Algorithm\ref{alg:swap} runs in time $O(n^{4})$.
\end{lem}
For the correctness of Algorithm \ref{alg:swap}, we have the following
lemma:
\begin{lem}
After the swap phase of Algorithm \ref{alg:mainA-simple-greedy},
there exist no ${\cal S}_{i}$ and ${\cal S}_{i'}$ with sub-edges
$j_{1},\,j_{2}\in{\cal S}_{i}$ and $j_{1}',\,j_{2}'\in{\cal S}_{i'}$,
such that $j_{1}<j_{1}'<j_{2}<j_{2}'$ holds.
\end{lem}
\begin{IEEEproof}
After the procession of ${\cal S}_{i}$, any ${\cal S}_{i'}\in{\cal S\setminus{\cal S}}_{i}$
must have all its sub-edges appear between the two edges $e_{j_{h}}\in{\cal S}_{i}$
and $e_{j_{h+1}}\in{\cal S}_{i}$ for some $h$. Then, after ${\cal S}_{i}$
is processed, Case (2) can not hold for any edge pair within ${\cal S}_{i}$
in any other latter iterations. Therefore, the algorithm guarantees
that one set contains no sub-edges of Case (2) at one iteration, and
hence after $n$ iterations, sub-edges of Case (2) are eliminated.
\end{IEEEproof}
The \emph{exchange} phase, invoked in Step 5 in Algorithm \ref{alg:rounding},
is to eliminate case (3) (i.e. $j_{1}<j_{1}'<j_{2}'<j_{2}$). The
key idea of the exchange is to move $j_{1}$ to the place exactly
before $j_{2}$ (or to move $j_{2}$ to the place exactly after $j_{1}$),
and then move the edges between $j_{1}$ and $j_{2}$ accordingly.
The choosing of movements (move $j_{1}$ to $j_{2}$, or $j_{2}$
to $j_{1}$) depends on the current \emph{offset }of the sub-edges
between $j_{1}$ and $j_{2}$, as well as the length of the edge $j_{1}$
and the length sum of the other edges in ${\cal S}_{i}\setminus\{j_{1}\}$,
where the \emph{offset} of an sub-edge $j$ is the distance from the
current position of $j$ to its original position. Formally, the \emph{exchange
}phase is as in Algorithm \ref{alg:exchange}.

\begin{algorithm}
Initially, each ${\cal S}_{i}$ contains a set of non-adjacent edges,
because the combining in Step 4 of Algorithm \ref{alg:mainLProundDMMSM}.
Assume that ${\cal E}=\{C_{1},\,\dots,\,C_{k}\}$ is the current set
of edges, which appear on the barrier from left to right in the order
of $C_{1},\,C_{2},\,\dots,$ $C_{k}$;

\enskip{}1: Set ${\cal O}:=\{o_{1},\,\dots,\,o_{k}\}$ and $o_{i}:=0$
for all $i$ initially; /{*}$o_{i}$ is the current movement offset
for $C_{i}$. {*}/

\enskip{}2: \textbf{While} true \textbf{do}

\enskip{}3: \quad{}Find the minimum $i$ such that there exists
$C_{i+\Delta}$ shares an identical ${\cal S}_{j_{i}}$ with $C_{i}$
for some $\Delta>1$;

\enskip{}4: \quad{}\textbf{If} no such $\Delta$ exists \textbf{then}

\enskip{}5: \quad{}\quad{}terminates;

\enskip{}6: \quad{}Find the minimum $\delta>1$ such that $C_{i+\delta}$
shares an identical ${\cal S}_{j_{i}}$ with $C_{i}$;

\enskip{}7: \quad{}Mover (${\cal E}$, ${\cal O}$, $i$, $i+\delta$);
/{*} Move $C_{i+\delta}$ or $C_{i}$, and the edges between them
accordingly. {*}/

\enskip{}8: Return ${\cal S}$.

\caption{\label{alg:exchange}Exchange(${\cal S}$).}
\end{algorithm}

In Algorithm \ref{alg:exchange}, the function Mover (${\cal E}$,
${\cal O}$, $i$, $i+\delta$) actually decides whether to move $C_{i+\delta}$
or $C_{i}$, according to which one of the two values $\vert C_{i}\vert-o_{i}$
and $\sum_{j:\,C_{j}\in{\cal S}_{j_{i}}\setminus\{C_{i}\}}\vert C_{j}\vert+o_{i}$
is larger. Intuitionally, without considering offsets, if we move
$C_{i}$ then the moving distance of $C_{l}$ between $C_{i}$ and
$C_{i+\delta}$ will be $\vert C_{i}\vert$; if move $C_{i+\delta}$,
the moving distance will be $\sum_{j:\,C_{j}\in{\cal S}_{j_{i}}\setminus\{C_{i}\}}\vert C_{j}\vert$
instead of $\vert C_{i+\delta}\vert$, since not only $C_{i+\delta}$
but every $C_{j}\in{\cal S}_{j_{i}}\setminus\{C_{i}\}$ will be moved
to adjacent to $C_{i}$. Then considering the offsets, we have the
criteria of moving $C_{i}$ or $C_{i+\delta}$. The moving algorithm
is as in Algorithm \ref{alg:mover}.

\begin{algorithm}
\enskip{}1: \textbf{If} $\vert C_{i}\vert-o_{i}\geq\sum_{j:\,C_{j}\in{\cal S}_{j_{i}}\setminus\{C_{i}\}}\vert C_{j}\vert+o_{i}$\textbf{
then} /{*}Move $C_{i+\delta}$ to the place adjacent to and in the
right side of $C_{i}$.{*}/

\enskip{}2: \quad{}Set ${\cal E}:={\cal E}\setminus\{C_{i+\delta}\}$
and update the numbering of the edges and vertices of ${\cal {\cal E}}$
accordingly;

\enskip{}3: \quad{}\textbf{For} $j=1$ to $\delta$ \textbf{do}

\enskip{}4: \quad{}\quad{}$x(v_{i+j}):=x(v_{i+j})+\vert C_{i+\delta}\vert$;

\enskip{}5: \quad{}\textbf{For} $j=1$ to $\delta-1$ \textbf{do
}/{*}Set the offset accordingly.{*}/

\enskip{}6: \quad{}\quad{}Set $o_{i+j}:=o_{i+j}+\vert C_{i+\delta}\vert$;

\enskip{}7: \textbf{Else} /{*}Move $C_{i+\delta}$. The case is similar
to line 1-8.{*}/

\enskip{}8: \quad{}Set ${\cal E}:={\cal E}\setminus\{C_{i}\}$ and
update the numbering of ${\cal {\cal E}}$ accordingly

\enskip{}9: \quad{}\textbf{For} $j=1$ to $\delta$ \textbf{do}

10: \quad{}\quad{}$x(v_{i+j}):=x(v_{i+j})-\vert C_{i}\vert$;

11: \quad{}\textbf{For} $j=1$ to $\delta-1$ \textbf{do}

12: \quad{}\quad{}Set $o_{i+j}:=o_{i+j}-\vert C_{i}\vert$;

\caption{\label{alg:mover}Mover (${\cal E}$, ${\cal O}$, $i$, $i+\delta$).}
\end{algorithm}

\begin{lem}
In Algorithm \ref{alg:rounding}, a sensor needs only at most $D+r_{max}$
movement to cover $C_{1},\,\dots,\,C_{k}$ when the given instance
is feasible wrt $D$.
\end{lem}
\begin{IEEEproof}
Clearly, after the swap phase sensor $i$ needs at most $D$ to any
edges of $S_{i}$. It remains to analysis the exchanging phase. We
will show that for a component $C_{i}$, its offset satisfies $-r_{max}\leq o_{i}\leq r_{max}$.

Let $o_{i_{1}}$ be the first non-zero value of $o_{i}$. Clearly
$-r_{max}\leq o_{i_{1}}\leq r_{max}$ holds, since $o_{i_{1}}$ is
actually $\min\{\vert C_{i_{1}}\vert,\,\vert C_{i_{1}+\delta_{1}}\vert\}$,
and $\vert C_{i_{1}}\vert+\vert C_{i_{1}+\delta_{1}}\vert\leq2r_{max}$,
where $\delta_{1}>1$ is minimum that $C_{i_{1}}$ and $C_{i_{1}+\delta_{1}}$
shares an identical ${\cal S}_{j}\in{\cal S}$. Then after the $t$th
times that $o_{i}$ changes, we have $o_{i_{t}}=\min\{\vert C_{i_{t}}\vert-o_{i_{t-1}},\,\vert C_{i_{t}+\delta_{t}}\vert+o_{i_{t-1}}\}$
according to Algorithm \ref{alg:mover}. So $o_{i_{t}}\leq\frac{(\vert C_{i_{t}}\vert-o_{i_{t-1}})+(\vert C_{i_{t}+\delta_{t}}\vert+o_{i_{t-1}})}{2}\leq r_{max}$.
On the other hand, from the induction hypothesis, $-r_{max}\leq o_{i_{t-1}}\leq r_{max}$
holds. So $o_{i_{t}}\geq-r_{max}$, since $\vert C_{i_{t}}\vert,\,\vert C_{i_{t}+\delta_{t}}\vert\geq0$.
\end{IEEEproof}
\begin{lem}
Algorithm \ref{alg:rounding} terminates in time $O(n^{7}L)$.
\end{lem}
\begin{IEEEproof}
The algorithm takes $O(n^{7}L)$ to solve LP1 by Karmakar's algorithm
\cite{karmarkar1984new}, since there are $O(n^{2})$ variables in
LP1. The \emph{swap} phase in the algorithm \ref{alg:rounding} takes
at most $O(n^{4})$ time as in Lemma \ref{lem:swaptime}, while the
\emph{exchange} phase iterates at most $O(k)=O(n^{2})$ times, each
iteration takes $O(k)=O(n^{2})$ time to run Mover (${\cal E}$, ${\cal O}$,
$i$, $i+\delta$). Other steps takes trivial time compared to the
above time, so the time complexity of the algorithm is $O(n^{7}L)$.
\end{IEEEproof}

\section{A Matching-Based Solution to Decision MMSM }

This subsection gives a pseudo polynomial algorithm for decision MMSM.
The key idea of the algorithm is to consider MMSM as DMMSM with uniform
edge length, where the barrier to cover is composed by $M$ edges
of length one. Then our algorithm is similar to the case in Section
3, excepting that we use maximum cardinality matching instead of fractional
cardinality matching to compute an initial solution. Using a similar
algorithm as in Subsection 3.3, we can round the initial solution,
i.e. the maximum cardinality matching, to a solution to MMSM.

To model a given instance of MMSM as maximum cardinality matching,
we will construct an equivalent bipartite graph $H=(U,\,V,\,E)$ in
which the vertex set $V$ corresponds to the sensors, the vertex set
$U$ corresponds to the edge set $\cup_{i}J_{i}$, where $J_{i}$
contains exactly the edges that can be completely sensed by sensor
$i$ within maximum movement $D$, and the edge set $E_{H}$ corresponds
to the coverage of the sensors to the edge of $\cup_{i}J_{i}$. Then
we check whether there exists a maximal cardinality matching with
size $M$ in $H$. If no such matching exists, the instance of DMMSM
is infeasible under maximum relocation distance $D$. Otherwise, similarly
as in Subsection 3.3, we can aggregate the vertices of $U$ that are
fractionally covered by an identical sensor, such that the sensors
can relocate within distance $D+r_{max}$ to cover all the edges.
The formal layout of the algorithm is as in Algorithm \ref{alg:mainMatchingbasedalgorithm.}.

\begin{algorithm}
\textbf{Input:} An instance of decision MMSM wrt a given $D$;

\textbf{Output: }$H=(U,\,V,\,E_{H})$.

\enskip{}1:\textbf{ For} each sensor $i$ \textbf{do }/{*}Compute
$J_{i}$ for each $i$. {*}/

\enskip{}2:\quad{}Compute $l_{i}$ and $g_{i}$ wrt $D$;

\enskip{}3:\quad{}Set $J_{i}=\left\{ j\vert j\in\left[\left\lceil l_{i}\right\rceil ,\,\left\lfloor g_{i}\right\rfloor \right]\right\} $;

\enskip{}4:\textbf{ For} $i=0$ to $M$ \textbf{do}

\enskip{}5: \quad{}Add a vertex $u_{i}$ to $U$;

\enskip{}6:\textbf{ For} each sensor $i$ \textbf{do}

\enskip{}7: \quad{}Add $2r_{i}$ vertices $\{v_{i,\,1},\,\dots,\,v_{i,\,2r_{i}}\}$
to $V$;

\enskip{}8:\textbf{ For} every pair of $u_{j}$ and $v_{i,\,l}$
\textbf{do}

\enskip{}9: \quad{}\textbf{If} $e_{j}\in J_{i}$ \textbf{then} Add
edge $(u_{j},\,v_{i,\,l})$ to $E_{H}$;

10: Compute a maximal cardinality matching ${\cal Y}$ for $H$;

11: \textbf{If} $\vert{\cal Y}\vert=M$ \textbf{then} return ``feasible'';

12: \textbf{Else }return ``infeasible''.

\caption{\label{alg:mainMatchingbasedalgorithm.}An matching-based approximation
algorithm for decision MMSM. }
\end{algorithm}

\begin{lem}
Let $R=\sum_{i\in{\cal I}}r_{i}$. Algorithm \ref{alg:mainMatchingbasedalgorithm.}
terminates in time $O\left(R^{2}\sqrt{\frac{M}{\log R}}\right)$.
\end{lem}
\begin{IEEEproof}
The maximal cardinality matching problem is known can be solved in
time $O(\vert V_{H}\vert\sqrt{\frac{\vert V_{H}\vert\vert E_{H}\vert}{\log\vert V_{H}\vert}})$,
where $\vert V_{H}\vert$ is the number of vertices and $\vert V_{H}\vert$
is number of edges. Following Algorithm \ref{alg:mainMatchingbasedalgorithm.},
$\vert V_{H}\vert=M+2R$ and $\vert E_{H}\vert\leq M\cdot2R$. So
the time needed to compute the matching as in Step 10 is actually
$O\left(R\sqrt{\frac{MR^{2}}{\log R}}\right)=O\left(R^{2}\sqrt{\frac{M}{\log R}}\right)$.
Other steps of the algorithm take trivial time compared to compute
the matching.
\end{IEEEproof}
\begin{thm}
\label{thm:improvedalgorithm}Let $D^{*}$ be the minimum movement
under which a given instance of MMSM is feasible. If Algorithm \ref{alg:mainMatchingbasedalgorithm.}
returns ``infeasible'', then $D<D^{*}$; Otherwise, the computed
matching ${\cal Y}$ can be transferred to a true solution to MMSM,
with a maximum relocation distance $D_{SOL}\leq D+r_{max}$.
\end{thm}
Firstly and apparently, if the DMMSM is feasible wrt a given $D$,
there must exist a matching with size $M$ in the corresponding graph
$H$. So if no such matching exists wrt $D$, then the instance DMMSM
must be infeasible wrt $D$. Secondly, similar to Algorithm \ref{alg:rounding},
the computed matching ${\cal Y}$ can be rounded to a true solution
to MMSM using \emph{swap} and \emph{exchange}.

\section{The Complete Algorithm for Solving MMSM}

The key idea of computing approximately a minimum $D$ is to use binary
search and call Algorithm \ref{alg:mainA-simple-greedy} (Or equivalently
Algorithm \ref{alg:mainLProundDMMSM} or Algorithm \ref{alg:mainMatchingbasedalgorithm.})
as a subroutine. Let $d_{i}$ be the minimum distance between sensor
$i$ and any point on the line barrier, and $d_{max}$ be the maximum
distance between the sensors and the barriers. Then clearly every
sensor of $\Gamma$ can cover any point of the barrier within movement
distance $d_{max}$. That is, within movement distance $d_{max}$
the line barrier can be covered successfully by the sensors of $\Gamma$;
Or the sensors in $\Gamma$ is not enough to cover the barrier. Then,
to find the minimum $D$ for MMSM, we need only to use binary search
within the range from 0 to $d_{max}$. Apparently, this takes at most
$O(\log(d_{max}))$ calls of Algorithm \ref{alg:mainA-simple-greedy}
(Or \ref{alg:mainLProundDMMSM}, \ref{alg:mainMatchingbasedalgorithm.})
to find the min-max feasible relocation distance bounded $D_{SOL}$.
Formally, the complete algorithm is as in Algorithm \ref{alg:greeedybasedThe-complete-algorithm1}.

\begin{algorithm}
\textbf{Input: }An instance of MMSM;

\textbf{Output:} The approximate min-max relocation distance $D_{SOL}$,
wrt which MMSM is feasible.

\enskip{}1: \textbf{If} $\sum_{i}2r_{i}<M$ \textbf{then}

\enskip{}2: \quad{}return ``infeasible'';

\enskip{}3: Compute $d_{max}=\max_{i\in[n]^{+},\,x_{l}\in[0,\,M]}\sqrt{(x_{i}-x_{l})^{2}+y_{i}^{2}}$;

\enskip{}4: Set $lower:=0$, $D:=upper:=d_{max}$; /{*} Clearly,
under maximum movement $upper$, the line barrier can be completely
covered. {*}/

\enskip{}5: \textbf{For} each sensor $i$ \textbf{do}

\enskip{}6: \quad{}Compute the leftmost position $l_{i}$ and the
rightmost position $g_{i}$ it can cover wrt $D$;

\enskip{}7: \quad{}Set $J_{i}=\{e_{j}\vert l_{i}\leq j\leq g_{i}\}$;

\enskip{}8: Call Algorithm \ref{alg:mainA-simple-greedy} to determine
whether the instance of MMSM is feasible under$D$;

\enskip{}9: \textbf{If} ``infeasible''\textbf{ then}

10: \quad{}Set $lower:=D$ and then $D=\frac{upper+D}{2}$;

11: \quad{}Go to Step 5;

12: \textbf{Else}

13: \quad{}\textbf{If} $upper-lower\leq1$ \textbf{then}

14: \quad{}\quad{}Return $D_{SOL}=D+2r_{max}$ ); /{*}The algorithm
terminates and outputs the solution.{*}/

15: \quad{}Set $upper:=D$, and then $D:=\frac{lower+D}{2}$ ;

16: \quad{}Go to Step 5.

\caption{\label{alg:greeedybasedThe-complete-algorithm1}An approximation algorithm
for MMSM. }
\end{algorithm}

From Theorem \ref{thm:greedyfinalthr}, we immediately have the time
complexity and ratio for the algorithm as follows:
\begin{lem}
Algorithm \ref{alg:greeedybasedThe-complete-algorithm1} terminates
in time $O(M\log d_{max})$, and output the relocation positions for
the sensors, within a maximum relocation distance $D+2r_{max}$.
\end{lem}
Note that, if using Algorithm \ref{alg:mainLProundDMMSM} instead
of Algorithm \ref{alg:mainA-simple-greedy} in Step 8, the runtime
and the maximum relocation distance of Algorithm \ref{alg:greeedybasedThe-complete-algorithm1}
will be $O(n^{7}L)$ and $D_{SOL}=D+r_{max}$, respectively.

\section{A Simple Factor-2 Approximation Algorithm for MMSM}

Following paper \cite{chen2013algorithms}, MMSM is solvable in time
$O(n^{2}\log n)$ if all the sensors are on the line containing the
barrier. Then a natural idea to solve 2D-MMSM is firstly to perpendicularly
move (some of) the sensors to the line barrier, and secondly solve
the consequent 1D-MMSM by using the algorithm in paper \cite{chen2013algorithms}.

Let $d_{p,\,i}$ be the perpendicular distance between sensor $i$
and the line barrier. Without loss of generality we assume that $0=d_{p,\,0}\leq d_{p,\,1}\leq\dots\leq d_{p,\,i}\leq\dots\leq d_{p,\,n}$,
where sensor $0$ is a virtual sensor with radii 0. Let $S(d_{p,\,i})=\{j\vert d_{p,\,j}\leq d_{p,\,i}\}$
be the set of sensors whose perpendicular distance to the line barrier
is not larger than $d_{p,\,i}$. Let $D_{h}(S(d_{p,\,i}))$ be the
maximum \emph{horizontal} relocation distance of the sensors in $S(d_{p,\,i})$
covering the barrier. Our algorithm will first simply compute $D_{h}(S(d_{p,\,i}))$
for every $i\in[n]^{+}$, and then select $\min\{d_{p,\,i}+D_{h}(S(d_{p,\,i}))\vert i\in[n]^{+}\}$
as the maximum relocation distance.
\begin{lem}
$\min\{d_{p,\,i}+D_{h}(S(d_{p,\,i}))\vert i\in[n]^{+}\}\leq2D^{*}$.
\end{lem}
\begin{IEEEproof}
For an optimum relocation solution to 2D-MMSM, assume that $d_{p,\,i^{*}}$
is the maximum perpendicular distance of the relocated sensors. Then
$d_{p,\,i^{*}}\leq D^{*}$, since sensor $i^{*}$ has to move at least
distance $d_{p,\,i^{*}}$ to cover the barrier. On the other hand,
apparently we have $D_{h}(S(d_{p,\,i^{*}}))\leq D^{*}$ i.e. the optimum
maximum horizontal relocation distance is not larger than $D^{*}$.
Therefore, we have $\min\{d_{p,\,i}+D_{h}(S(d_{p,\,i}))\vert i\in[n]^{+}\}\leq d_{p,\,i^{*}}+D_{h}(S(d_{p,\,i^{*}}))\leq2D^{*}$.
\end{IEEEproof}
Clearly, the above naive algorithm has to run the 1D-MMSM algorithm
for $O(n)$ times to compute $\min\{d_{p,\,i}+D_{h}(S(d_{p,\,i}))\vert i\in[n]^{+}\}$.
Hence, it runs in time $O(n^{3}\log n)$. Note that the binary search
cannot be immediately applied here, since $f(i)=d_{p,\,i}+D_{h}(S(d_{p,\,i}))$
is neither monotonously increasing nor monotonously decreasing on
$i$. Anyhow, we will give an improve algorithm in which the number
of times of solving 1D-MMSM is improved to $O(\log n)$. We will show
the ratio of the improved algorithm remains two by giving further
observations.

The key idea of our improved algorithm is, instead of finding an $i$
such that $d_{p,\,i}+D_{h}(S(d_{p,\,i}))$ attends minimum, to find
an $i_{alt}$, such that $D_{h}(S(d_{p,\,i_{alt}})\leq d_{p,\,i_{alt}}$
and $D_{h}(S(d_{p,\,i_{alt}-1})\geq d_{p,\,i_{alt}-1}$ both hold.
Note that such $i_{alt}$ can be found with solving the 1D-MMSM problem
only for $O(\log n)$ times, via a binary search in which a set of
values $[lwr,\,upp]$, $lwr\leq i_{alt}\leq upp$, is maintained.
The formal layout of the algorithm is as in Algorithm \ref{alg:factor2apprx}.

\begin{algorithm}
\textbf{Input: }An instance of MMSM, in which w.l.o.g. assume that
$y_{1}\leq\dots\leq y_{n}$;

\textbf{Output:} $i_{alt}$.

\enskip{}1: Set $upp:=n$, $j:=n$ and $lwr:=1$;

\enskip{}2: \textbf{While} $upp-lwr>1$ \textbf{do}

\enskip{}3: \quad{}Set $S(d_{p,\,j})=\{i\vert i\leq j,\,i\in S\}$,
and set the position of $i$ therein to $(x_{i},\,0)$;

\enskip{}4: \quad{}Solve the 1D-MMSM problem with respect to $S(d_{p,\,j})$
and $S(d_{p,\,j-1})$, respectively, using the algorithm as in \cite{chen2013algorithms};

\quad{}\quad{}/{*} Obtain $D_{h}(S(d_{p,\,i_{alt}-1})$ and $D_{h}(S(d_{p,\,i_{alt}})$.{*}/

\enskip{}5:\quad{}\textbf{If} $D_{h}(S(d_{p,\,j})\geq d_{p,\,j}$
\textbf{then }/{*}The current value of $j$ is too small. {*}/

\enskip{}6:\quad{}\quad{}Set $lwr:=j$, and $j:=\frac{upp+lwr}{2}$;

\enskip{}7:\quad{}\textbf{Else}

\enskip{}8:\quad{}\quad{}Set $upp:=j$, and $j:=\frac{upp+lwr}{2}$;

\enskip{}9: Return $i_{alt}:=j$.

\caption{\label{alg:factor2apprx}A factor-2 approximation algorithm for 2D-MMSM. }
\end{algorithm}

The basic observation of our algorithm is that $f(i)=D_{h}(S(d_{p,\,i}))$
will not increase while $i$ increases.
\begin{prop}
\label{prop:monotonelydcrease}$f(i)=D_{h}(S(d_{p,\,i}))$ is monotonously
decreasing on $i$.
\end{prop}
The correctness of the above proposition immediately follows from
the fact that $S(d_{p,\,i})\supseteq S(d_{p,\,i-1})$. Then the performance
guarantee of the algorithm is as in the following lemma:
\begin{lem}
$\min\{d_{p,\,i_{alt}}+D_{h}(S(d_{p,\,i_{alt}}),\,d_{p,\,i_{alt}-1}+D_{h}(S(d_{p,\,i_{alt}-1})\}\leq2D^{*}$.
\end{lem}
\begin{IEEEproof}
Assume that $d_{p,\,i^{*}}$ is the maximum perpendicular distance
of the relocated sensors in an optimum relocation solution to 2D-MMSM.
Then we show that $\min\{d_{p,\,i_{alt}}+D_{h}(S(d_{p,\,i_{alt}}),\,d_{p,\,i_{alt}-1}+D_{h}(S(d_{p,\,i_{alt}-1})\}\leq2D^{*}$
holds for either $i^{*}\geq i_{alt}$ or $i^{*}<i_{alt}$.

\begin{enumerate}
\item $i^{*}\geq i_{alt}$:

Obviously, we have $d_{p,\,i_{alt}}\leq d_{p,\,i^{*}}\leq D^{*}$.
Then since $D_{h}(S(d_{p,\,i_{alt}})\leq d_{p,\,i_{alt}}$, we have
$d_{p,\,i_{alt}}+D_{h}(S(d_{p,\,i_{alt}})\leq2D^{*}$.
\item $i^{*}<i_{alt}$:

Since $i^{*}\leq i_{alt}$ and that $f(i)=D_{h}(S(d_{p,\,i})$ is
monotonously decreasing on $i$, we have $D_{h}(S(d_{p,\,i_{alt}-1})\leq D_{h}(S(d_{p,\,i^{*}})\leq D^{*}$.
Then since $D_{h}(S(d_{p,\,i_{alt}-1})\geq d_{p,\,i_{alt}-1}$ according
to the algorithm, we have $d_{p,\,i_{alt}-1}\leq D^{*}$. Therefore,
$d_{p,\,i_{alt}-1}+D_{h}(S(d_{p,\,i_{alt}-1})\leq2D^{*}$ holds.
\end{enumerate}
\end{IEEEproof}
Since Algorithm \ref{alg:factor2apprx} iterates the \emph{while-loop}
for at most $O(\log n)$ times, we have the following theorem:
\begin{thm}
MMSM admits a factor-2 approximation algorithm with runtime $O(n^{2}(\log n)^{2})$.
\end{thm}

\section{Conclusion}

This paper developed three algorithms for MMSM via solving decision
MMSM, which are respectively with runtime $O(n\log n \log d_{max})$, $O(n^{7}L\log d_{max})$
and $O\left(R^{2}\sqrt{\frac{M}{\log R}}\log d_{max}\right)$, and
maximum relocation distance $d(OPT)+2r_{max}$, $d(OPT)+r_{max}$
and $d(OPT)+r_{max}$, where $n$ is the number of sensors, $L$ is
the length of input, $M$ is the length of the barrier, $d(OPT)$
is the maximum relocation distance in an optimum solution to MMSM,
$d_{max}$ is the maximum distance between the sensors and the barriers,
and $r_{max}=\max_{i}\{r_{i}\}$ is the maximum sensing radii of the
sensors. We proved the performance guarantee for the first algorithm
by giving a sufficient condition to check the feasibility of an instance
of decision MMSM, and for the second (and hence the third) algorithm
by rounding up an optimum fractional solution against the according
LP relaxation, to a real solution of DMMSM. To the best of our knowledge,
our method of rounding up a fractional LP solution is the first to
round an LP solution by aggregation, and has the potential to be applied
to solve other problems. In addition, we developed a factor-2 approximation
by extending a previous result in paper \cite{chen2013algorithms}.
Consequently, the performance of our first three algorithms can be
improved when $r_{max}>d(OPT)$. We note that our proposed algorithms
can only work for MMSM with only one barrier, and are currently investigating
approximation algorithms for MMSM with multiple barriers.

\bibliographystyle{plain}
\bibliography{ref}

\end{document}